\documentclass[preprint,prb,showpacs,floatfix,superscriptaddress]{revtex4-1}

\usepackage[linktocpage,bookmarksopen,bookmarksnumbered]{hyperref}
\usepackage[utf8]{inputenc} 
\usepackage{dcolumn}
\usepackage{graphicx}
\usepackage{amsfonts}
\usepackage{amsmath,bm,amssymb}
\usepackage{braket} 

\begin{document}

\title{Charge density functional plus $U$ calculation of lacunar spinel GaM$_4$Se$_8$ (M = Nb, Mo, Ta, and W)}

\author{Hyunggeun Lee}
\author{Min Yong Jeong}
\author{Jae-Hoon Sim}
\author{Hongkee Yoon}
\author{Siheon Ryee}
\author{Myung Joon Han} \email{mj.han@kaist.ac.kr}

\affiliation{Department of Physics, KAIST, 291 Daehak-ro, Yuseong-gu, Daejeon 34141, Republic of Korea}

\begin{abstract}

Charge density functional plus $U$ calculations are carried out to examine the validity of molecular $J_\text{eff}$=1/2 and 3/2 state in lacunar spinel GaM$_4$X$_8$ (M = Nb, Mo, Ta, and W). With LDA (spin-unpolarized local density approximation)$+U$, which has recently been suggested as the more desirable choice than LSDA (local spin density approximation)$+U$, we examine the band structure in comparison with the previous prediction based on the spin-polarized version of functional and with the prototypical $J_\text{eff}$=1/2 material Sr$_2$IrO$_4$. It is found that the previously suggested $J_\text{eff}$=1/2 and 3/2 band characters remain valid still in LDA$+U$ calculations while the use of charge-only density causes some minor differences. Our result provides the further support for the novel molecular $J_\text{eff}$ state in this series of materials, which can hopefully motivate the future exploration toward its verification and the further search for new functionalities.
\end{abstract}

\maketitle

\section{Introduction}
A series of `lacunar spinel' compounds, GaM$_4$X$_8$ (M = V, Nb, Mo, and Ta; X = S, Se, and Te), have attracted great attention due to their interesting physical properties and promising material characteristics. For example, multiferroicity has been observed in GaV$_4$S$_8$ and GaV$_4$Se$_8$ carrying a great potential for memory device applications \cite{ruff_multiferroicity_2015, widmann_multiferroic_2017, reschke_optical_2017, ruff_polar_2017}. In GaTa$_4$Se$_8$, resistive switching phenomena which can be used for resistive random access memory (RRAM) have been reported \cite{dubost_resistive_2013}. In the case of M=Nb and Ta, the insulator-to-metal transition followed by superconducting transition is known to occur by applying pressure \cite{abd-elmeguid_transition_2004, pocha_crystal_2005}. Their intriguing low temperature behaviors in susceptibility and specific heat measurement \cite{yaich_nouveaux_1984, pocha_crystal_2005, kawamoto_frustrated_2016} can possibly be related to the unconventional superconductivity. Further, novel `molecular $J_\text{eff}$' ground states have been suggested recently. According to the first-principles band calculations, the molecular $J_\text{eff}$=1/2 and 3/2 state are realized in the case of M = Mo, W, and Nb, Ta, respectively, due to the crucial role of spin-orbit interaction while this effect has been ignored in earlier studies \cite{kim_spin-orbital_2014}. For GaTa$_4$Se$_8$, the novel $J_\text{eff}$=3/2 ground state has been verified by resonant inelastic x-ray scattering (RIXS) experiment combined with theoretical calculations \cite{jeong_direct_2017}.

One important next step is therefore to study the other materials (i.e., GaNb$_4$Se$_8$, GaMo$_4$Se$_8$, and GaW$_4$Se$_8$) and to confirm their characteristic molecular $J_\text{eff}$ states, which can provide a new exciting playground in search for the novel quantum phenomena \cite{jackeli_mott_2009, chaloupka_kitaev-heisenberg_2010, watanabe_microscopic_2010, wang_twisted_2011, kim_magnetic_2012, dey_spin-liquid_2012, kimchi_kitaev-heisenberg_2014, kim_fermi_2014, chun_direct_2015, kim_observation_2016}. While the similar type of experiments such as RIXS and RXMS (resonant x-ray magnetic scattering) can be utilized \cite{kim_phase-sensitive_2009, jeong_direct_2017}, only available at this moment is the band structure prediction \cite{kim_spin-orbital_2014}. On the one hand, the successful verification of $J_\text{eff}$=3/2 for the case of M = Ta \cite{jeong_direct_2017} supports the reliability of the previous theoretical prediction \cite{kim_spin-orbital_2014}. On the other, a series of recent DFT (density functional theory)+$U$ studies require the further investigation. According to recent careful studies, the use of charge-only density functional (such as LDA and GGA (spin-unpolarized generalized gradient approximation)) is highly desirable for DFT+$U$ type of calculation rather than the use of spin density functional (such as LSDA  and SGGA (spin-polarized GGA)) \cite{chen_density_2015, park_density_2015, chen_spin-density_2016, ryee_effect_2018, ryee_comparative_2018,comment}. In LSDA+$U$ or SGGA+$U$ scheme, the intrinsic Stoner type exchange interactions can likely cause the unphysical magnetic behaviors through the uncontrolled competition and double counting with the interaction term like Hund exchange. This feature has been noticed in some case studies \cite{chen_density_2015, park_density_2015, chen_spin-density_2016} and then further analyzed in a formal way \cite{ryee_effect_2018, ryee_comparative_2018}. Since the previous band structure prediction of molecular $J_\text{eff}$ states has been based on SGGA+$U$ calculation with the functional form suggested by Dudarev et al. \cite{dudarev_electron-energy-loss_1998, han_$mathrmon$_2006, kim_spin-orbital_2014}, it is necessary to confirm the validity of it.

In this paper, we performed LDA+$U$ calculations and confirmed the robustness of $J_\text{eff}$ band structure for the $4d$ and $5d$ lacunar spinels. It is found that  $J_\text{eff}=1/2$ and 3/2 Mott ground states are well maintained in the reasonably large range of Hubbard $U$ and Hund $J$ parameters. By introducing a new quantity $b_\text{avg}$, which is designed to measure the $J_\text{eff}=1/2$ and 3/2 band separation, we present the quantitative argument in comparison to the previous SGGA+$U$ results and the prototype $J_\text{eff}=1/2$ material, Sr$_2$IrO$_4$. By confirming the previous theoretical prediction, our result hopefully motivates the further research toward the verification of these exotic quantum states and the search for the  new functionality.

\section{Computation details}
Band structure calculations were carried out using `OpenMX' software package which is based on the linear combination of pseudo-atomic orbital (LCPAO) formalism \cite{ozaki_variationally_2003}. LDA+$U$ calculations were performed  with the (so-called) fully localized limit (FLL) functional form \cite{anisimov_density-functional_1993, solovyev_corrected_1994, czyzyk_local-density_1994, liechtenstein_density-functional_1995}. The spin-orbit coupling (SOC) was treated within the fully relativistic $j$-dependent pseudopotential \cite{macdonald_spin-polarised_1983, bachelet_pseudopotentials_1982, theurich_self-consistent_2001}. We adopted the energy cutoff of 400 Rydberg for real space grid and $12 \times 12 \times 12$ $k$-points for the primitive unit cell. For the reasonable values of Hubbard $U$ and Hund $J$, we took the previous cRPA (constrained random-phase approximation) results for each transition-metal element as our reference \cite{sasioglu_effective_2011}.  Considering that our lacunar spinels are all insulating, we used the 25\% larger values of $U$ than the cRPA results for elements since the Mott gap is opened at around this value for GaNb$_4$Se$_8$ and GaTa$_4$Se$_8$:
$U=$ 3.4, 4.5, 3.0, and 4.4 eV ~ for GaNb$_4$Se$_8$, GaMo$_4$Se$_8$, GaTa$_4$Se$_8$ and GaW$_4$Se$_8$, respectively. The crystal structures were optimized, and for the band structure, we present the 2-formula-unit cell results with the antiferromagnetic inter-cluster order.

In order to discuss the robustness of $J_\text{eff}$ nature in a quantitative way, we introduce a new parameter:
\begin{align}
b_{n,{\boldsymbol{k}}}&=\frac{\left|2P^{n,{\boldsymbol{k}}}_{1/2}-P^{n,{\boldsymbol{k}}}_{3/2}\right|}{2P^{n,{\boldsymbol{k}}}_{1/2}+P^{n,{\boldsymbol{k}}}_{3/2}}.
\end{align}
Here
\begin{align}
P^{n,{\boldsymbol{k}}}_{1/2}=&\left|\braket{\frac{1}{2}\ +\frac{1}{2}|\psi_{n,{\boldsymbol{k}}}}\right|^{2}+\left|\braket{\frac{1}{2}\ -\frac{1}{2}|\psi_{n,{\boldsymbol{k}}}}\right|^{2} \\
{\text{and}} \nonumber\\
P^{n,{\boldsymbol{k}}}_{3/2}=&\left|\braket{\frac{3}{2}\ +\frac{3}{2}|\psi_{n,{\boldsymbol{k}}}}\right|^{2}+\left|\braket{\frac{3}{2}\ +\frac{1}{2}|\psi_{n,{\boldsymbol{k}}}}\right|^{2}\nonumber\\
&+\left|\braket{\frac{3}{2}\ -\frac{1}{2}|\psi_{n,{\boldsymbol{k}}}}\right|^{2}+\left|\braket{\frac{3}{2}\ -\frac{3}{2}|\psi_{n,{\boldsymbol{k}}}}\right|^{2}\nonumber,
\end{align}
where $n$ and ${\boldsymbol{k}}$ represent the band index and the momentum, respectively. Also, the atomic $J_\text{eff}$ states are written as $\bra{J_{\text{eff}}\ m_{J_{\text{eff}}}}$. The  $b_{n,{\boldsymbol{k}}}$ quantifies the ratio between $J_\text{eff}=1/2$ and $J_\text{eff}=3/2$ with a factor 2 which represents the statistical ratio, 2:1, for the ideal $t_\text{2g}$ or molecular $t_2$ states. Thus, ideally, $b_{n,{\boldsymbol{k}}}$ becomes 0 if the given state is a purely atomic $t_\text{2g}$ or molecular $t_2$ state (i.e., well identified just by $\ket{xy\ \pm\sigma}$, $\ket{xz\ \pm\sigma}$, and $\ket{yz\ \pm\sigma}$) while it becomes 1.0 when the eigenstate is identical with the pure $J_\text{eff}$ states. Now the separation of $J_\text{eff}=1/2$ and $J_\text{eff}=3/2$ bands of a given material can be represented by taking the average for the entire space:\\

\begin{figure}[t!]
	\centering
	\includegraphics[width=0.75\linewidth]{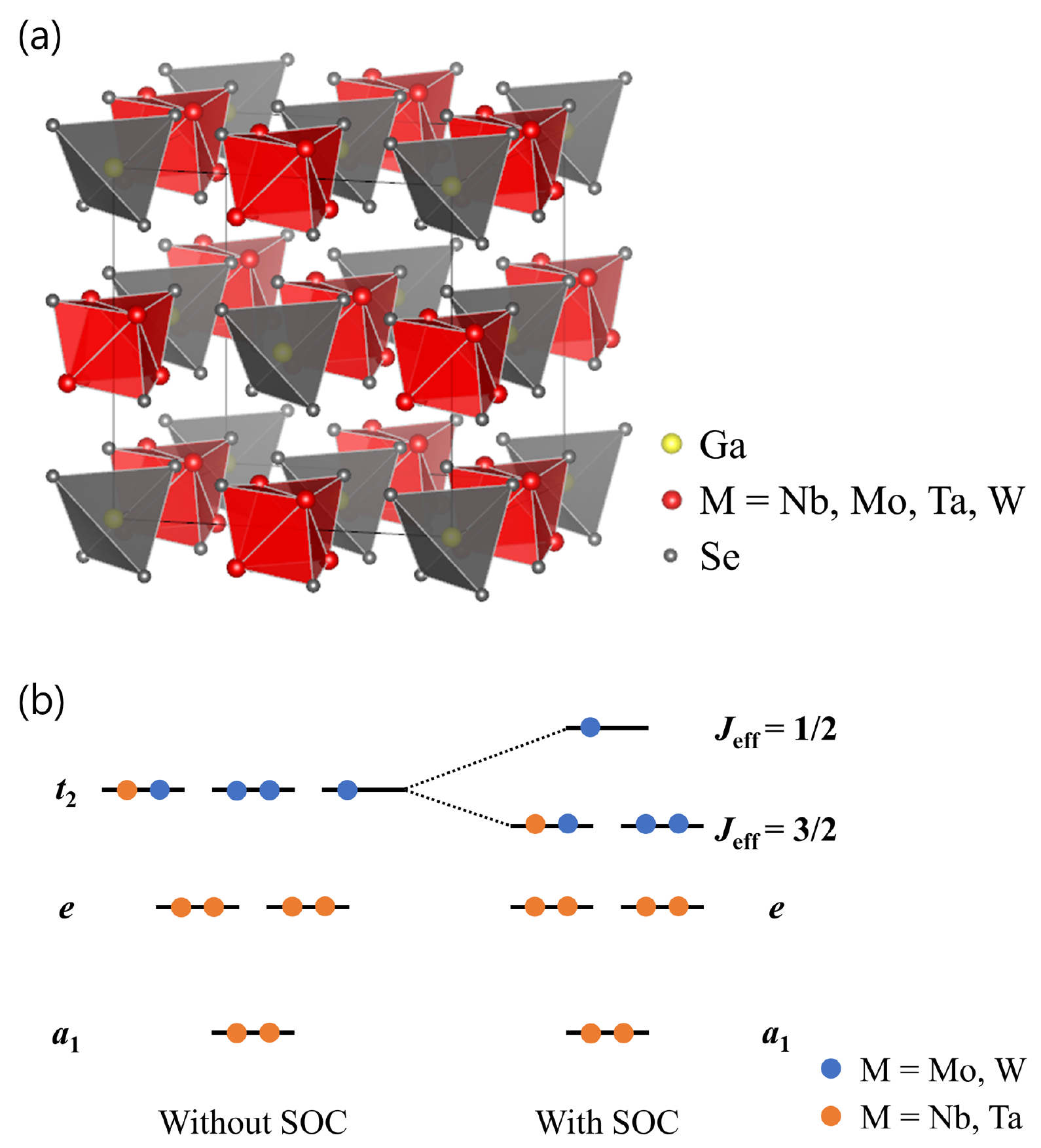}
	\caption{(a) Crystal structure of GaM$_4$Se$_8$ (cubic F$\bar{4}3m$). The yellow, red, and grey spheres represent Ga, M (= Nb, Mo, Ta, and W) and Se atoms respectively. GaM$_4$Se$_8$ is composed of M$_4$Se$_4$ (red) and GaSe$_4$ (grey) clusters. (b) Schematic energy level diagrams for M$_4$ cluster with and without SOC. Without SOC, $t_2$ molecular orbital states are 6-fold degenerate. SOC splits them  into 2-fold $J_\text{eff}=1/2$ and 4-fold degenerate $J_\text{eff}=3/2$ states.}
	\label{fig:fig1}
\end{figure}

\begin{equation}
b_\text{avg}=\frac{\sum_{\boldsymbol{k}}{\sum_{n}{\left(P^{n,{\boldsymbol{k}}}_{1/2}+P^{n,{\boldsymbol{k}}}_{3/2}\right)b_{n,{\boldsymbol{k}}}}}}{\sum_{\boldsymbol{k}}{\sum_{n}{\left(P^{n,{\boldsymbol{k}}}_{1/2}+P^{n,{\boldsymbol{k}}}_{3/2}\right)}}}.
\end{equation}

This value therefore provides a single number with which the $J_\text{eff}$ nature of band structure can be expressed. As an example, let us consider the prototypical $J_\text{eff}$=1/2 material Sr$_2$IrO$_4$. The calculated $b_\text{avg}$ based on the `(so-called) Dudarev functional \cite{dudarev_electron-energy-loss_1998}' with $U=2.0$ yields $b_\text{avg}=0.486$. This is the case for the original calculation result by Kim et al.~\cite{kim_novel_2008}. If we performed the calculation with the SOC turned off, $b_\text{avg}$ becomes zero. Here we also performed LDA+$U$ calculation for Sr$_2$IrO$_4$ with the 25\% larger value of $U$ than the cRPA value for elemental Ir. The result remains same; $b_\text{avg}=0.487$. Here it should be noted that $b_\text{avg}$ depends on the degree of $d$ orbitals hybridization with other orbitals (e.g., oxygen or chalcogen $p$), the local structure, and crystal field splitting, etc. Therefore the interpretation and the comparison of the absolute values of $b_\text{avg}$ need to be careful.

\section{Results and discussion}

\begin{figure*}[t!]
	\centering
	\includegraphics[width=1.0\linewidth]{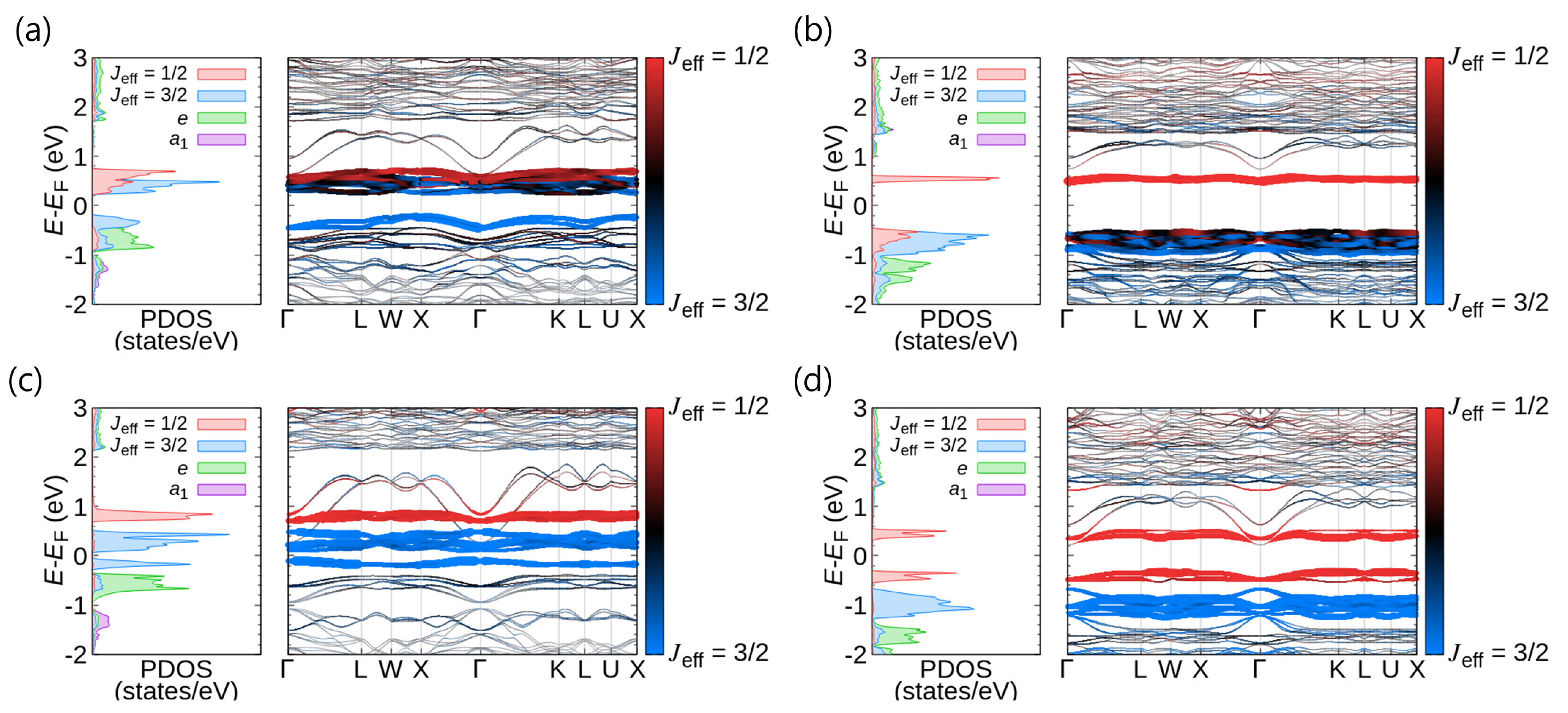}
	\caption{The calculated $J_\text{eff}$-projected DOS and the band dispersions for (a) GaNb$_4$Se$_8$ ($U=3.4$ eV, $J=0.45$ eV), (b) GaMo$_4$Se$_8$ ($U=4.5$ eV, $J=0.5$ eV), (c) GaTa$_4$Se$_8$ ($U=3.0$ eV, $J=0.4$ eV) and (d) GaW$_4$Se$_8$ ($U=4.4$ eV, $J=0.45$ eV). The red, blue, green and violet colors in DOS represent the $J_\text{eff}=1/2$, $J_\text{eff}=3/2$, $e$ and $a_1$ character, respectively. In the band dispersion, the $J_\text{eff}=1/2$ (red) and $3/2$ (blue) character are represented by the line thickness.}
	\label{fig:fig2}
\end{figure*}

The crystal structure of lacunar spinel GaM$_4$Se$_8$ (space group F$\bar{4}$3m) can be understood as deduced from the spinel, GaM$_2$Se$_4$, with half-deficient Ga atoms \cite{francois_structural_1991, pocha_crystal_2005}. A characteristic feature is that the well-defined molecular units of M$_4$Se$_4$ and GaSe$_4$ are arranged to be NaCl structure as shown in Fig.~1(a). The 12-fold M-M bonding complex is split into 6-fold $t_2$, 4-fold $e$ and 2-fold degenerate $a_1$ states due to $T_d$ molecular symmetry \cite{johrendt_crystal_1998, pocha_electronic_2000, nakamura_structural_2005, guiot_control_2011, ta_phuoc_optical_2013, kim_spin-orbital_2014, camjayi_first-order_2014} as shown in Fig.~1(b).  The electronic structure near the Fermi level is governed by molecular $t_2$ states which are derived from the atomic $t_\text{2g}$ orbitals of transition metals \cite{johrendt_crystal_1998, pocha_crystal_2005, ta_phuoc_optical_2013, kim_spin-orbital_2014, camjayi_first-order_2014}. It is noted that the molecular $t_2$ orbitals have the same symmetry with atomic $t_\text{2g}$, and the SOC leads them to split into the `effective’ angular momentum $J_\text{eff}=1/2$ doublet and $J_\text{eff}=3/2$ quartet \cite{kim_spin-orbital_2014}. Depending on the number of valence electrons, GaMo$_4$Se$_8$ and GaW$_4$Se$_8$ have $J_\text{eff}=1/2$ while GaNb$_4$Se$_8$ and GaTa$_4$Se$_8$ carry $J_\text{eff}=3/2$ moment; see Fig.~1(b). These `molecular’ $J_\text{eff}$ ground states were first predicted by band structure calculation \cite{kim_spin-orbital_2014}, and the case for $J_\text{eff}=3/2$ has recently been confirmed for M = Ta \cite{jeong_direct_2017}.

Fig.~2 presents the projected density of states (PDOS; left panels) and the fat band dispersion (right panels) obtained by LDA+$U$ calculations; (a) GaNb$_4$Se$_8$, (b) GaMo$_4$Se$_8$, (c) GaTa$_4$Se$_8$, and (d) GaW$_4$Se$_8$. First of all, we note that the characteristic molecular $J_\text{eff}$ states are well maintained as in the previous calculation of SGGA+$U$ functional \cite{kim_spin-orbital_2014}. The upper/lower Hubbard bands are predominantly of $J_\text{eff}=1/2$ and $J_\text{eff}=3/2$ character for GaW$_4$Se$_8$ and GaTa$_4$Se$_8$, respectively; see Fig.~2(c) and (d). For $4d$ materials, the mixture between the two $J_\text{eff}$ states is noticed in the upper and lower Hubbard part for GaNb$_4$Se$_8$ and GaMo$_4$Se$_8$, respectively (see Fig.2(a) and (b)), which is a comparable feature with the case of Sr$_2$IrO$_4$ \cite{arita_ab_2012}.

In particular, for GaTa$_4$Se$_8$, the higher-lying $J_\text{eff}=1/2$ peak is located at around +0.8 eV ~and well identified (i.e., having a negligible mixture with other states). This feature together with the lower-lying `$e$' bands at $<$$-$0.3 eV ~ (see the green-colored PDOS in Fig.2(c)) is mainly responsible for the novel quantum interference observed in a recent RIXS experiment which is the first direct experimental evidence for $J_\text{eff}=3/2$ moment in a real material \cite{jeong_direct_2017}. Thus, our current result provides the further confirmation of the characteristic electronic structure of GaTa$_4$Se$_8$ and other lacunar spinels by using the charge-only LDA+$U$ calculations which have recently been suggested as the more desirable functional choice than SGGA+$U$ \cite{chen_spin-density_2016, ryee_effect_2018}. 

\begin{figure}
	\centering
	\includegraphics[width=0.75\linewidth]{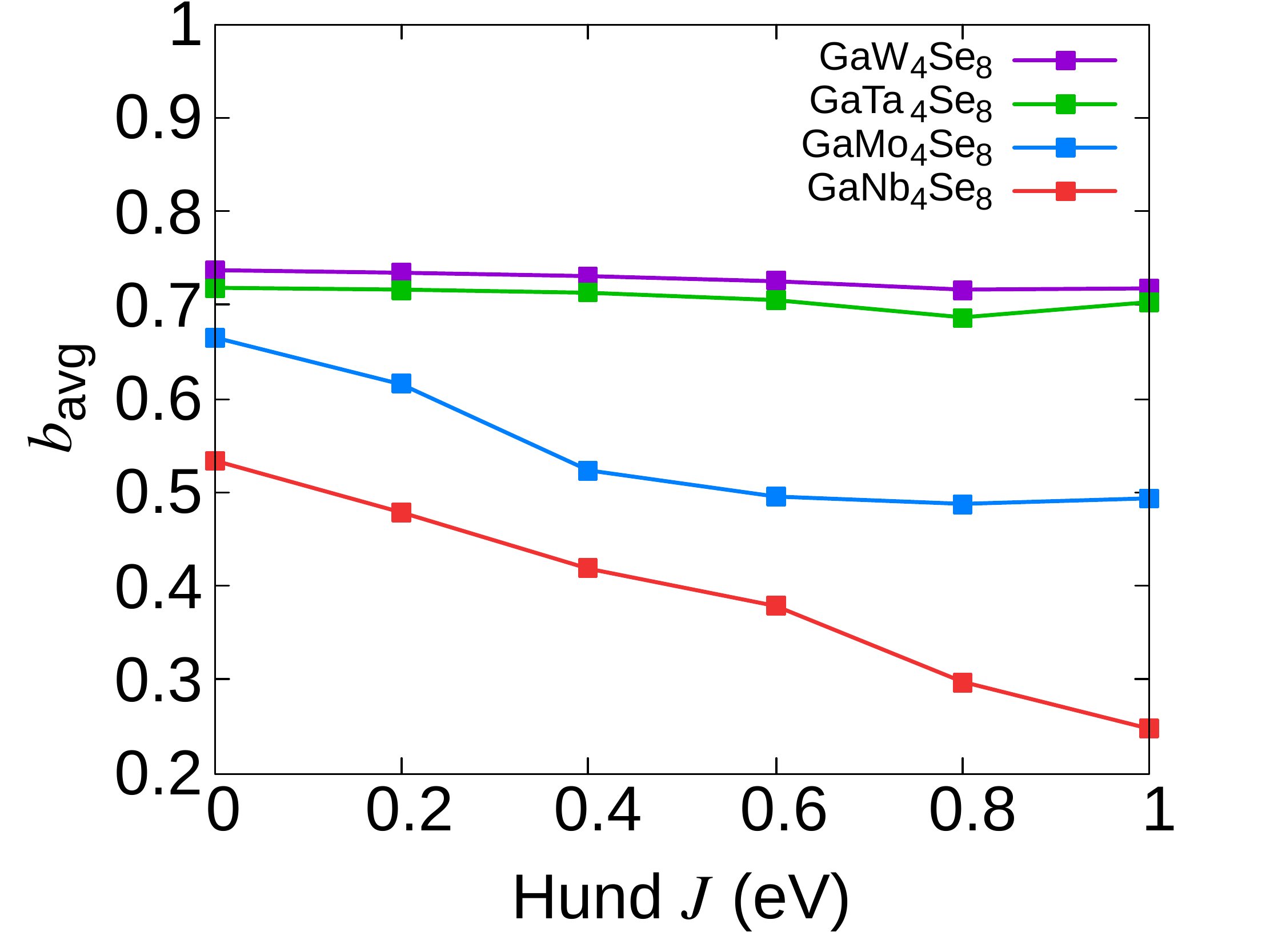}
	\caption{The calculated $b_\text{avg}$ of four lacunar spinel compounds as a function of Hund $J$. The violet, green, blue, and red color represents the results of GaW$_4$Se$_8$, GaTa$_4$Se$_8$, GaMo$_4$Se$_8$, and GaNb$_4$Se$_8$, respectively.}
	\label{fig:fig3}
\end{figure}

The differences in the two calculated band structure, namely by LDA+$U$ and LSDA+$U$, are relatively minor. For all four compounds, the calculated band gaps are smaller in LDA+$U$ than LSDA+$U$ by about 0.1--0.2 eV ~at the same interaction parameters of $U$ and $J$. It implies that the larger $U$ is required to open the Mott gap in LDA+$U$. The band separation between $J_\text{eff}=1/2$ and $J_\text{eff}=3/2$ is slightly more pronounced in LDA+$U$ results. This feature can also be seen by comparing the calculated $b_\text{avg}$ which will be discussed in the below.

One obvious limitation of Dudarev formalism is its inability to calculate Hund-$J$-dependent electronic structure \cite{dudarev_electron-energy-loss_1998,kim_spin-orbital_2014}. In order to check the robustness of $J_\text{eff}$ band character in lacunar spinels, we performed the calculations as a function of Hund $J$; see Fig.~3. The results show that the degree of separation of $J_\text{eff}=1/2$ and $J_\text{eff}=3/2$ bands as represented by $b_\text{avg}$ is well maintained in the wide range of $J$, especially for $5d$ compounds. The calculated $b_\text{avg}$ for GaTa$_4$Se$_8$ and GaW$_4$Se$_8$ are quite large ($b_\text{avg} > 0.68$) in the entire range of Hund $J$ considered in this study (see the violet and green line). Due to the smaller SOC, on the other hand, the calculated $b_\text{avg}$ for $4d$ compounds depends more severely on Hund $J$. In the large $J$ limit, the calculated $b_\text{avg}$ becomes as small as $\sim$0.49 and $\sim$0.25 for GaMo$_4$Se$_8$ and GaNb$_4$Se$_8$, respectively (see the blue and red line). In the reasonable range of Hund $J \approx 0.5$ eV, $b_\text{avg}$ is about 0.5 for both Nb and Mo cases which is comparable with the value of Sr$_2$IrO$_4$. While $b_\text{avg}$ is just a simple measure of the degree of separation of $J_\text{eff}=1/2$ and $J_\text{eff}=3/2$ bands based on the calculated electronic structure, our calculation clearly supports the robustness of the molecular $J_\text{eff}$ character of lacunar spinels.

In order to further check if the $J_\text{eff}$ band character remains valid, we calculated $b_\text{avg}$ as a function of $U$ with the fixed $J$ to the cRPA values: $J=$0.45, 0.50, 0.40, and 0.45 eV ~ for GaNb$_4$Se$_8$, GaMo$_4$Se$_8$, GaTa$_4$Se$_8$, and GaW$_4$Se$_8$. In a wide range of $U$ value from 2.0 to 4.5 eV, we found that $b_\text{avg}$ does not change much. For $5d$ compounds of M = Ta and W, the calculated $b_\text{avg}$ remains well above $b_\text{avg}=0.6$ and mainly close to 0.7 for $U\leq$ 4.0 eV. In the case of GaMo$_4$Se$_8$, the calculated $b_\text{avg}$ remains not smaller than 0.5. For M = Nb, the calculated $b_\text{avg}$ is gradually reduced from 0.43 at $U=3.0$ eV ~to 0.33 at $U=4.5$ eV, which is noticeably smaller than the other three compounds. While the quantified $b_\text{avg}$ is certainly smaller in $4d$ materials, we think that GaNb$_4$Se$_8$ can also be well identified as a molecular $J_\text{eff}=3/2$ material especially considering that its material properties are quite similar with those of GaTa$_4$Se$_8$ including the insulator-to-metal transition and superconductivity under pressure \cite{abd-elmeguid_transition_2004}. It would be an interesting experimental challenge to confirm this exotic ground state for M = Nb just as in the recent report on GaTa$_4$Se$_8$ \cite{jeong_direct_2017}.

\section{Conclusion}
With LDA+$U$ calculations, we confirm the previous theoretical prediction based on SGGA$+U$ for the molecular $J_\text{eff}$ band structures in $4d$/$5d$ lacunar spinels. By introducing a new parameter, $b_{\text{avg}}$, we performed the quantitative examination of $J_\text{eff}$ band separation as a function of $J$ which was not feasible in the previous study. It is found that $5d$ compounds have the quite robust $J_\text{eff}$ band character while both $4d$ and $5d$ materials have well-identified $J_\text{eff}$ feature at around the realistic $J$ values. This $J_\text{eff}$ nature is also quite well maintained in the reasonable range of $U$. Our results provide the solid guidance for future study of this materials by strengthening the theoretical prediction of the novel material characteristic. In particular, the detailed magnetic property at low temperature and under pressure need to be further identified and understood, which can also elucidate the nature of superconductivity found in $J_\text{eff}$=3/2 materials.

\acknowledgments
This work was supported by Basic Science Research Program through the National Research Foundation of Korea (NRF) funded by the Ministry of Education (2018R1A2B2005204) and Creative Materials Discovery Program through the NRF funded by Ministry of Science and ICT (2018M3D1A1058754).

\bibliography{ref.bib}

\begin{thebibliography}{48}%
\makeatletter
\providecommand \@ifxundefined [1]{%
 \@ifx{#1\undefined}
}%
\providecommand \@ifnum [1]{%
 \ifnum #1\expandafter \@firstoftwo
 \else \expandafter \@secondoftwo
 \fi
}%
\providecommand \@ifx [1]{%
 \ifx #1\expandafter \@firstoftwo
 \else \expandafter \@secondoftwo
 \fi
}%
\providecommand \natexlab [1]{#1}%
\providecommand \enquote  [1]{``#1''}%
\providecommand \bibnamefont  [1]{#1}%
\providecommand \bibfnamefont [1]{#1}%
\providecommand \citenamefont [1]{#1}%
\providecommand \href@noop [0]{\@secondoftwo}%
\providecommand \href [0]{\begingroup \@sanitize@url \@href}%
\providecommand \@href[1]{\@@startlink{#1}\@@href}%
\providecommand \@@href[1]{\endgroup#1\@@endlink}%
\providecommand \@sanitize@url [0]{\catcode `\\12\catcode `\$12\catcode
  `\&12\catcode `\#12\catcode `\^12\catcode `\_12\catcode `\%12\relax}%
\providecommand \@@startlink[1]{}%
\providecommand \@@endlink[0]{}%
\providecommand \url  [0]{\begingroup\@sanitize@url \@url }%
\providecommand \@url [1]{\endgroup\@href {#1}{\urlprefix }}%
\providecommand \urlprefix  [0]{URL }%
\providecommand \Eprint [0]{\href }%
\providecommand \doibase [0]{http://dx.doi.org/}%
\providecommand \selectlanguage [0]{\@gobble}%
\providecommand \bibinfo  [0]{\@secondoftwo}%
\providecommand \bibfield  [0]{\@secondoftwo}%
\providecommand \translation [1]{[#1]}%
\providecommand \BibitemOpen [0]{}%
\providecommand \bibitemStop [0]{}%
\providecommand \bibitemNoStop [0]{.\EOS\space}%
\providecommand \EOS [0]{\spacefactor3000\relax}%
\providecommand \BibitemShut  [1]{\csname bibitem#1\endcsname}%
\let\auto@bib@innerbib\@empty
\bibitem [{\citenamefont {Ruff}\ \emph {et~al.}(2015)\citenamefont {Ruff},
  \citenamefont {Widmann}, \citenamefont {Lunkenheimer}, \citenamefont
  {Tsurkan}, \citenamefont {Bordács}, \citenamefont {Kézsmárki},\ and\
  \citenamefont {Loidl}}]{ruff_multiferroicity_2015}%
  \BibitemOpen
  \bibfield  {author} {\bibinfo {author} {\bibfnamefont {E.}~\bibnamefont
  {Ruff}}, \bibinfo {author} {\bibfnamefont {S.}~\bibnamefont {Widmann}},
  \bibinfo {author} {\bibfnamefont {P.}~\bibnamefont {Lunkenheimer}}, \bibinfo
  {author} {\bibfnamefont {V.}~\bibnamefont {Tsurkan}}, \bibinfo {author}
  {\bibfnamefont {S.}~\bibnamefont {Bordács}}, \bibinfo {author}
  {\bibfnamefont {I.}~\bibnamefont {Kézsmárki}}, \ and\ \bibinfo {author}
  {\bibfnamefont {A.}~\bibnamefont {Loidl}},\ }\href {\doibase
  10.1126/sciadv.1500916} {\bibfield  {journal} {\bibinfo  {journal} {Sci.
  Adv.}\ }\textbf {\bibinfo {volume} {1}},\ \bibinfo {pages} {e1500916}
  (\bibinfo {year} {2015})}\BibitemShut {NoStop}%
\bibitem [{\citenamefont {Widmann}\ \emph {et~al.}(2017)\citenamefont
  {Widmann}, \citenamefont {Ruff}, \citenamefont {Günther}, \citenamefont
  {Nidda}, \citenamefont {Lunkenheimer}, \citenamefont {Tsurkan}, \citenamefont
  {Bordács}, \citenamefont {Kézsmárki},\ and\ \citenamefont
  {Loidl}}]{widmann_multiferroic_2017}%
  \BibitemOpen
  \bibfield  {author} {\bibinfo {author} {\bibfnamefont {S.}~\bibnamefont
  {Widmann}}, \bibinfo {author} {\bibfnamefont {E.}~\bibnamefont {Ruff}},
  \bibinfo {author} {\bibfnamefont {A.}~\bibnamefont {Günther}}, \bibinfo
  {author} {\bibfnamefont {H.-A. K.~v.}\ \bibnamefont {Nidda}}, \bibinfo
  {author} {\bibfnamefont {P.}~\bibnamefont {Lunkenheimer}}, \bibinfo {author}
  {\bibfnamefont {V.}~\bibnamefont {Tsurkan}}, \bibinfo {author} {\bibfnamefont
  {S.}~\bibnamefont {Bordács}}, \bibinfo {author} {\bibfnamefont
  {I.}~\bibnamefont {Kézsmárki}}, \ and\ \bibinfo {author} {\bibfnamefont
  {A.}~\bibnamefont {Loidl}},\ }\href {\doibase 10.1080/14786435.2016.1253885}
  {\bibfield  {journal} {\bibinfo  {journal} {Philos. Mag.}\ }\textbf {\bibinfo
  {volume} {97}},\ \bibinfo {pages} {3428} (\bibinfo {year}
  {2017})}\BibitemShut {NoStop}%
\bibitem [{\citenamefont {Reschke}\ \emph {et~al.}(2017)\citenamefont
  {Reschke}, \citenamefont {Mayr}, \citenamefont {Wang}, \citenamefont
  {Lunkenheimer}, \citenamefont {Li}, \citenamefont {Szaller}, \citenamefont
  {Bordács}, \citenamefont {Kézsmárki}, \citenamefont {Tsurkan},\ and\
  \citenamefont {Loidl}}]{reschke_optical_2017}%
  \BibitemOpen
  \bibfield  {author} {\bibinfo {author} {\bibfnamefont {S.}~\bibnamefont
  {Reschke}}, \bibinfo {author} {\bibfnamefont {F.}~\bibnamefont {Mayr}},
  \bibinfo {author} {\bibfnamefont {Z.}~\bibnamefont {Wang}}, \bibinfo {author}
  {\bibfnamefont {P.}~\bibnamefont {Lunkenheimer}}, \bibinfo {author}
  {\bibfnamefont {W.}~\bibnamefont {Li}}, \bibinfo {author} {\bibfnamefont
  {D.}~\bibnamefont {Szaller}}, \bibinfo {author} {\bibfnamefont
  {S.}~\bibnamefont {Bordács}}, \bibinfo {author} {\bibfnamefont
  {I.}~\bibnamefont {Kézsmárki}}, \bibinfo {author} {\bibfnamefont
  {V.}~\bibnamefont {Tsurkan}}, \ and\ \bibinfo {author} {\bibfnamefont
  {A.}~\bibnamefont {Loidl}},\ }\href {\doibase 10.1103/PhysRevB.96.144302}
  {\bibfield  {journal} {\bibinfo  {journal} {Phys. Rev. B}\ }\textbf {\bibinfo
  {volume} {96}},\ \bibinfo {pages} {144302} (\bibinfo {year}
  {2017})}\BibitemShut {NoStop}%
\bibitem [{\citenamefont {Ruff}\ \emph {et~al.}(2017)\citenamefont {Ruff},
  \citenamefont {Butykai}, \citenamefont {Geirhos}, \citenamefont {Widmann},
  \citenamefont {Tsurkan}, \citenamefont {Stefanet}, \citenamefont
  {Kézsmárki}, \citenamefont {Loidl},\ and\ \citenamefont
  {Lunkenheimer}}]{ruff_polar_2017}%
  \BibitemOpen
  \bibfield  {author} {\bibinfo {author} {\bibfnamefont {E.}~\bibnamefont
  {Ruff}}, \bibinfo {author} {\bibfnamefont {A.}~\bibnamefont {Butykai}},
  \bibinfo {author} {\bibfnamefont {K.}~\bibnamefont {Geirhos}}, \bibinfo
  {author} {\bibfnamefont {S.}~\bibnamefont {Widmann}}, \bibinfo {author}
  {\bibfnamefont {V.}~\bibnamefont {Tsurkan}}, \bibinfo {author} {\bibfnamefont
  {E.}~\bibnamefont {Stefanet}}, \bibinfo {author} {\bibfnamefont
  {I.}~\bibnamefont {Kézsmárki}}, \bibinfo {author} {\bibfnamefont
  {A.}~\bibnamefont {Loidl}}, \ and\ \bibinfo {author} {\bibfnamefont
  {P.}~\bibnamefont {Lunkenheimer}},\ }\href {\doibase
  10.1103/PhysRevB.96.165119} {\bibfield  {journal} {\bibinfo  {journal} {Phys.
  Rev. B}\ }\textbf {\bibinfo {volume} {96}},\ \bibinfo {pages} {165119}
  (\bibinfo {year} {2017})}\BibitemShut {NoStop}%
\bibitem [{\citenamefont {Dubost}\ \emph {et~al.}(2013)\citenamefont {Dubost},
  \citenamefont {Cren}, \citenamefont {Vaju}, \citenamefont {Cario},
  \citenamefont {Corraze}, \citenamefont {Janod}, \citenamefont
  {Debontridder},\ and\ \citenamefont {Roditchev}}]{dubost_resistive_2013}%
  \BibitemOpen
  \bibfield  {author} {\bibinfo {author} {\bibfnamefont {V.}~\bibnamefont
  {Dubost}}, \bibinfo {author} {\bibfnamefont {T.}~\bibnamefont {Cren}},
  \bibinfo {author} {\bibfnamefont {C.}~\bibnamefont {Vaju}}, \bibinfo {author}
  {\bibfnamefont {L.}~\bibnamefont {Cario}}, \bibinfo {author} {\bibfnamefont
  {B.}~\bibnamefont {Corraze}}, \bibinfo {author} {\bibfnamefont
  {E.}~\bibnamefont {Janod}}, \bibinfo {author} {\bibfnamefont
  {F.}~\bibnamefont {Debontridder}}, \ and\ \bibinfo {author} {\bibfnamefont
  {D.}~\bibnamefont {Roditchev}},\ }\href {\doibase 10.1021/nl401510p}
  {\bibfield  {journal} {\bibinfo  {journal} {Nano Lett.}\ }\textbf {\bibinfo
  {volume} {13}},\ \bibinfo {pages} {3648} (\bibinfo {year}
  {2013})}\BibitemShut {NoStop}%
\bibitem [{\citenamefont {Abd-Elmeguid}\ \emph {et~al.}(2004)\citenamefont
  {Abd-Elmeguid}, \citenamefont {Ni}, \citenamefont {Khomskii}, \citenamefont
  {Pocha}, \citenamefont {Johrendt}, \citenamefont {Wang},\ and\ \citenamefont
  {Syassen}}]{abd-elmeguid_transition_2004}%
  \BibitemOpen
  \bibfield  {author} {\bibinfo {author} {\bibfnamefont {M.~M.}\ \bibnamefont
  {Abd-Elmeguid}}, \bibinfo {author} {\bibfnamefont {B.}~\bibnamefont {Ni}},
  \bibinfo {author} {\bibfnamefont {D.~I.}\ \bibnamefont {Khomskii}}, \bibinfo
  {author} {\bibfnamefont {R.}~\bibnamefont {Pocha}}, \bibinfo {author}
  {\bibfnamefont {D.}~\bibnamefont {Johrendt}}, \bibinfo {author}
  {\bibfnamefont {X.}~\bibnamefont {Wang}}, \ and\ \bibinfo {author}
  {\bibfnamefont {K.}~\bibnamefont {Syassen}},\ }\href {\doibase
  10.1103/PhysRevLett.93.126403} {\bibfield  {journal} {\bibinfo  {journal}
  {Phys. Rev. Lett.}\ }\textbf {\bibinfo {volume} {93}},\ \bibinfo {pages}
  {126403} (\bibinfo {year} {2004})}\BibitemShut {NoStop}%
\bibitem [{\citenamefont {Pocha}\ \emph {et~al.}(2005)\citenamefont {Pocha},
  \citenamefont {Johrendt}, \citenamefont {Ni},\ and\ \citenamefont
  {Abd-Elmeguid}}]{pocha_crystal_2005}%
  \BibitemOpen
  \bibfield  {author} {\bibinfo {author} {\bibfnamefont {R.}~\bibnamefont
  {Pocha}}, \bibinfo {author} {\bibfnamefont {D.}~\bibnamefont {Johrendt}},
  \bibinfo {author} {\bibfnamefont {B.}~\bibnamefont {Ni}}, \ and\ \bibinfo
  {author} {\bibfnamefont {M.~M.}\ \bibnamefont {Abd-Elmeguid}},\ }\href
  {\doibase 10.1021/ja050243x} {\bibfield  {journal} {\bibinfo  {journal} {J.
  Am. Chem. Soc.}\ }\textbf {\bibinfo {volume} {127}},\ \bibinfo {pages} {8732}
  (\bibinfo {year} {2005})}\BibitemShut {NoStop}%
\bibitem [{\citenamefont {Yaich}\ \emph {et~al.}(1984)\citenamefont {Yaich},
  \citenamefont {Jegaden}, \citenamefont {Potel}, \citenamefont {Sergent},
  \citenamefont {Rastogi},\ and\ \citenamefont
  {Tournier}}]{yaich_nouveaux_1984}%
  \BibitemOpen
  \bibfield  {author} {\bibinfo {author} {\bibfnamefont {H.~B.}\ \bibnamefont
  {Yaich}}, \bibinfo {author} {\bibfnamefont {J.~C.}\ \bibnamefont {Jegaden}},
  \bibinfo {author} {\bibfnamefont {M.}~\bibnamefont {Potel}}, \bibinfo
  {author} {\bibfnamefont {M.}~\bibnamefont {Sergent}}, \bibinfo {author}
  {\bibfnamefont {A.~K.}\ \bibnamefont {Rastogi}}, \ and\ \bibinfo {author}
  {\bibfnamefont {R.}~\bibnamefont {Tournier}},\ }\href {\doibase
  10.1016/0022-5088(84)90384-9} {\bibfield  {journal} {\bibinfo  {journal} {J.
  Less-Common Met.}\ }\textbf {\bibinfo {volume} {102}},\ \bibinfo {pages} {9}
  (\bibinfo {year} {1984})}\BibitemShut {NoStop}%
\bibitem [{\citenamefont {Kawamoto}\ \emph {et~al.}(2016)\citenamefont
  {Kawamoto}, \citenamefont {Higo}, \citenamefont {Tomita}, \citenamefont
  {Suzuki}, \citenamefont {Tian}, \citenamefont {Mochitzuki}, \citenamefont
  {Matsuo}, \citenamefont {Kindo},\ and\ \citenamefont
  {Nakatsuji}}]{kawamoto_frustrated_2016}%
  \BibitemOpen
  \bibfield  {author} {\bibinfo {author} {\bibfnamefont {S.}~\bibnamefont
  {Kawamoto}}, \bibinfo {author} {\bibfnamefont {T.}~\bibnamefont {Higo}},
  \bibinfo {author} {\bibfnamefont {T.}~\bibnamefont {Tomita}}, \bibinfo
  {author} {\bibfnamefont {S.}~\bibnamefont {Suzuki}}, \bibinfo {author}
  {\bibfnamefont {Z.~M.}\ \bibnamefont {Tian}}, \bibinfo {author}
  {\bibfnamefont {K.}~\bibnamefont {Mochitzuki}}, \bibinfo {author}
  {\bibfnamefont {A.}~\bibnamefont {Matsuo}}, \bibinfo {author} {\bibfnamefont
  {K.}~\bibnamefont {Kindo}}, \ and\ \bibinfo {author} {\bibfnamefont
  {S.}~\bibnamefont {Nakatsuji}},\ }\href {\doibase
  10.1088/1742-6596/683/1/012025} {\bibfield  {journal} {\bibinfo  {journal}
  {J. Phys. Conf. Ser.}\ }\textbf {\bibinfo {volume} {683}},\ \bibinfo {pages}
  {012025} (\bibinfo {year} {2016})}\BibitemShut {NoStop}%
\bibitem [{\citenamefont {Kim}\ \emph {et~al.}(2014{\natexlab{a}})\citenamefont
  {Kim}, \citenamefont {Im}, \citenamefont {Han},\ and\ \citenamefont
  {Jin}}]{kim_spin-orbital_2014}%
  \BibitemOpen
  \bibfield  {author} {\bibinfo {author} {\bibfnamefont {H.-S.}\ \bibnamefont
  {Kim}}, \bibinfo {author} {\bibfnamefont {J.}~\bibnamefont {Im}}, \bibinfo
  {author} {\bibfnamefont {M.~J.}\ \bibnamefont {Han}}, \ and\ \bibinfo
  {author} {\bibfnamefont {H.}~\bibnamefont {Jin}},\ }\href {\doibase
  10.1038/ncomms4988} {\bibfield  {journal} {\bibinfo  {journal} {Nat.
  Commun.}\ }\textbf {\bibinfo {volume} {5}},\ \bibinfo {pages} {3988}
  (\bibinfo {year} {2014}{\natexlab{a}})}\BibitemShut {NoStop}%
\bibitem [{\citenamefont {Jeong}\ \emph {et~al.}(2017)\citenamefont {Jeong},
  \citenamefont {Chang}, \citenamefont {Kim}, \citenamefont {Sim},
  \citenamefont {Said}, \citenamefont {Casa}, \citenamefont {Gog},
  \citenamefont {Janod}, \citenamefont {Cario}, \citenamefont {Yunoki},
  \citenamefont {Han},\ and\ \citenamefont {Kim}}]{jeong_direct_2017}%
  \BibitemOpen
  \bibfield  {author} {\bibinfo {author} {\bibfnamefont {M.~Y.}\ \bibnamefont
  {Jeong}}, \bibinfo {author} {\bibfnamefont {S.~H.}\ \bibnamefont {Chang}},
  \bibinfo {author} {\bibfnamefont {B.~H.}\ \bibnamefont {Kim}}, \bibinfo
  {author} {\bibfnamefont {J.-H.}\ \bibnamefont {Sim}}, \bibinfo {author}
  {\bibfnamefont {A.}~\bibnamefont {Said}}, \bibinfo {author} {\bibfnamefont
  {D.}~\bibnamefont {Casa}}, \bibinfo {author} {\bibfnamefont {T.}~\bibnamefont
  {Gog}}, \bibinfo {author} {\bibfnamefont {E.}~\bibnamefont {Janod}}, \bibinfo
  {author} {\bibfnamefont {L.}~\bibnamefont {Cario}}, \bibinfo {author}
  {\bibfnamefont {S.}~\bibnamefont {Yunoki}}, \bibinfo {author} {\bibfnamefont
  {M.~J.}\ \bibnamefont {Han}}, \ and\ \bibinfo {author} {\bibfnamefont
  {J.}~\bibnamefont {Kim}},\ }\href {\doibase 10.1038/s41467-017-00841-9}
  {\bibfield  {journal} {\bibinfo  {journal} {Nat. Commun.}\ }\textbf {\bibinfo
  {volume} {8}},\ \bibinfo {pages} {782} (\bibinfo {year} {2017})}\BibitemShut
  {NoStop}%
\bibitem [{\citenamefont {Jackeli}\ and\ \citenamefont
  {Khaliullin}(2009)}]{jackeli_mott_2009}%
  \BibitemOpen
  \bibfield  {author} {\bibinfo {author} {\bibfnamefont {G.}~\bibnamefont
  {Jackeli}}\ and\ \bibinfo {author} {\bibfnamefont {G.}~\bibnamefont
  {Khaliullin}},\ }\href {\doibase 10.1103/PhysRevLett.102.017205} {\bibfield
  {journal} {\bibinfo  {journal} {Phys. Rev. Lett.}\ }\textbf {\bibinfo
  {volume} {102}},\ \bibinfo {pages} {017205} (\bibinfo {year}
  {2009})}\BibitemShut {NoStop}%
\bibitem [{\citenamefont {Chaloupka}\ \emph {et~al.}(2010)\citenamefont
  {Chaloupka}, \citenamefont {Jackeli},\ and\ \citenamefont
  {Khaliullin}}]{chaloupka_kitaev-heisenberg_2010}%
  \BibitemOpen
  \bibfield  {author} {\bibinfo {author} {\bibfnamefont {J.}~\bibnamefont
  {Chaloupka}}, \bibinfo {author} {\bibfnamefont {G.}~\bibnamefont {Jackeli}},
  \ and\ \bibinfo {author} {\bibfnamefont {G.}~\bibnamefont {Khaliullin}},\
  }\href {\doibase 10.1103/PhysRevLett.105.027204} {\bibfield  {journal}
  {\bibinfo  {journal} {Phys. Rev. Lett.}\ }\textbf {\bibinfo {volume} {105}},\
  \bibinfo {pages} {027204} (\bibinfo {year} {2010})}\BibitemShut {NoStop}%
\bibitem [{\citenamefont {Watanabe}\ \emph {et~al.}(2010)\citenamefont
  {Watanabe}, \citenamefont {Shirakawa},\ and\ \citenamefont
  {Yunoki}}]{watanabe_microscopic_2010}%
  \BibitemOpen
  \bibfield  {author} {\bibinfo {author} {\bibfnamefont {H.}~\bibnamefont
  {Watanabe}}, \bibinfo {author} {\bibfnamefont {T.}~\bibnamefont {Shirakawa}},
  \ and\ \bibinfo {author} {\bibfnamefont {S.}~\bibnamefont {Yunoki}},\ }\href
  {\doibase 10.1103/PhysRevLett.105.216410} {\bibfield  {journal} {\bibinfo
  {journal} {Phys. Rev. Lett.}\ }\textbf {\bibinfo {volume} {105}},\ \bibinfo
  {pages} {216410} (\bibinfo {year} {2010})}\BibitemShut {NoStop}%
\bibitem [{\citenamefont {Wang}\ and\ \citenamefont
  {Senthil}(2011)}]{wang_twisted_2011}%
  \BibitemOpen
  \bibfield  {author} {\bibinfo {author} {\bibfnamefont {F.}~\bibnamefont
  {Wang}}\ and\ \bibinfo {author} {\bibfnamefont {T.}~\bibnamefont {Senthil}},\
  }\href {\doibase 10.1103/PhysRevLett.106.136402} {\bibfield  {journal}
  {\bibinfo  {journal} {Phys. Rev. Lett.}\ }\textbf {\bibinfo {volume} {106}},\
  \bibinfo {pages} {136402} (\bibinfo {year} {2011})}\BibitemShut {NoStop}%
\bibitem [{\citenamefont {Kim}\ \emph {et~al.}(2012)\citenamefont {Kim},
  \citenamefont {Casa}, \citenamefont {Upton}, \citenamefont {Gog},
  \citenamefont {Kim}, \citenamefont {Mitchell}, \citenamefont {van
  Veenendaal}, \citenamefont {Daghofer}, \citenamefont {van~den Brink},
  \citenamefont {Khaliullin},\ and\ \citenamefont {Kim}}]{kim_magnetic_2012}%
  \BibitemOpen
  \bibfield  {author} {\bibinfo {author} {\bibfnamefont {J.}~\bibnamefont
  {Kim}}, \bibinfo {author} {\bibfnamefont {D.}~\bibnamefont {Casa}}, \bibinfo
  {author} {\bibfnamefont {M.~H.}\ \bibnamefont {Upton}}, \bibinfo {author}
  {\bibfnamefont {T.}~\bibnamefont {Gog}}, \bibinfo {author} {\bibfnamefont
  {Y.-J.}\ \bibnamefont {Kim}}, \bibinfo {author} {\bibfnamefont {J.~F.}\
  \bibnamefont {Mitchell}}, \bibinfo {author} {\bibfnamefont {M.}~\bibnamefont
  {van Veenendaal}}, \bibinfo {author} {\bibfnamefont {M.}~\bibnamefont
  {Daghofer}}, \bibinfo {author} {\bibfnamefont {J.}~\bibnamefont {van~den
  Brink}}, \bibinfo {author} {\bibfnamefont {G.}~\bibnamefont {Khaliullin}}, \
  and\ \bibinfo {author} {\bibfnamefont {B.~J.}\ \bibnamefont {Kim}},\ }\href
  {\doibase 10.1103/PhysRevLett.108.177003} {\bibfield  {journal} {\bibinfo
  {journal} {Phys. Rev. Lett.}\ }\textbf {\bibinfo {volume} {108}},\ \bibinfo
  {pages} {177003} (\bibinfo {year} {2012})}\BibitemShut {NoStop}%
\bibitem [{\citenamefont {Dey}\ \emph {et~al.}(2012)\citenamefont {Dey},
  \citenamefont {Mahajan}, \citenamefont {Khuntia}, \citenamefont {Baenitz},
  \citenamefont {Koteswararao},\ and\ \citenamefont
  {Chou}}]{dey_spin-liquid_2012}%
  \BibitemOpen
  \bibfield  {author} {\bibinfo {author} {\bibfnamefont {T.}~\bibnamefont
  {Dey}}, \bibinfo {author} {\bibfnamefont {A.~V.}\ \bibnamefont {Mahajan}},
  \bibinfo {author} {\bibfnamefont {P.}~\bibnamefont {Khuntia}}, \bibinfo
  {author} {\bibfnamefont {M.}~\bibnamefont {Baenitz}}, \bibinfo {author}
  {\bibfnamefont {B.}~\bibnamefont {Koteswararao}}, \ and\ \bibinfo {author}
  {\bibfnamefont {F.~C.}\ \bibnamefont {Chou}},\ }\href {\doibase
  10.1103/PhysRevB.86.140405} {\bibfield  {journal} {\bibinfo  {journal} {Phys.
  Rev. B}\ }\textbf {\bibinfo {volume} {86}},\ \bibinfo {pages} {140405}
  (\bibinfo {year} {2012})}\BibitemShut {NoStop}%
\bibitem [{\citenamefont {Kimchi}\ and\ \citenamefont
  {Vishwanath}(2014)}]{kimchi_kitaev-heisenberg_2014}%
  \BibitemOpen
  \bibfield  {author} {\bibinfo {author} {\bibfnamefont {I.}~\bibnamefont
  {Kimchi}}\ and\ \bibinfo {author} {\bibfnamefont {A.}~\bibnamefont
  {Vishwanath}},\ }\href {\doibase 10.1103/PhysRevB.89.014414} {\bibfield
  {journal} {\bibinfo  {journal} {Phys. Rev. B}\ }\textbf {\bibinfo {volume}
  {89}},\ \bibinfo {pages} {014414} (\bibinfo {year} {2014})}\BibitemShut
  {NoStop}%
\bibitem [{\citenamefont {Kim}\ \emph {et~al.}(2014{\natexlab{b}})\citenamefont
  {Kim}, \citenamefont {Krupin}, \citenamefont {Denlinger}, \citenamefont
  {Bostwick}, \citenamefont {Rotenberg}, \citenamefont {Zhao}, \citenamefont
  {Mitchell}, \citenamefont {Allen},\ and\ \citenamefont
  {Kim}}]{kim_fermi_2014}%
  \BibitemOpen
  \bibfield  {author} {\bibinfo {author} {\bibfnamefont {Y.~K.}\ \bibnamefont
  {Kim}}, \bibinfo {author} {\bibfnamefont {O.}~\bibnamefont {Krupin}},
  \bibinfo {author} {\bibfnamefont {J.~D.}\ \bibnamefont {Denlinger}}, \bibinfo
  {author} {\bibfnamefont {A.}~\bibnamefont {Bostwick}}, \bibinfo {author}
  {\bibfnamefont {E.}~\bibnamefont {Rotenberg}}, \bibinfo {author}
  {\bibfnamefont {Q.}~\bibnamefont {Zhao}}, \bibinfo {author} {\bibfnamefont
  {J.~F.}\ \bibnamefont {Mitchell}}, \bibinfo {author} {\bibfnamefont {J.~W.}\
  \bibnamefont {Allen}}, \ and\ \bibinfo {author} {\bibfnamefont {B.~J.}\
  \bibnamefont {Kim}},\ }\href {\doibase 10.1126/science.1251151} {\bibfield
  {journal} {\bibinfo  {journal} {Science}\ }\textbf {\bibinfo {volume}
  {345}},\ \bibinfo {pages} {187} (\bibinfo {year}
  {2014}{\natexlab{b}})}\BibitemShut {NoStop}%
\bibitem [{\citenamefont {Chun}\ \emph {et~al.}(2015)\citenamefont {Chun},
  \citenamefont {Kim}, \citenamefont {Kim}, \citenamefont {Zheng},
  \citenamefont {Stoumpos}, \citenamefont {Malliakas}, \citenamefont
  {Mitchell}, \citenamefont {Mehlawat}, \citenamefont {Singh}, \citenamefont
  {Choi}, \citenamefont {Gog}, \citenamefont {Al-Zein}, \citenamefont {Sala},
  \citenamefont {Krisch}, \citenamefont {Chaloupka}, \citenamefont {Jackeli},
  \citenamefont {Khaliullin},\ and\ \citenamefont {Kim}}]{chun_direct_2015}%
  \BibitemOpen
  \bibfield  {author} {\bibinfo {author} {\bibfnamefont {S.~H.}\ \bibnamefont
  {Chun}}, \bibinfo {author} {\bibfnamefont {J.-W.}\ \bibnamefont {Kim}},
  \bibinfo {author} {\bibfnamefont {J.}~\bibnamefont {Kim}}, \bibinfo {author}
  {\bibfnamefont {H.}~\bibnamefont {Zheng}}, \bibinfo {author} {\bibfnamefont
  {C.~C.}\ \bibnamefont {Stoumpos}}, \bibinfo {author} {\bibfnamefont {C.~D.}\
  \bibnamefont {Malliakas}}, \bibinfo {author} {\bibfnamefont {J.~F.}\
  \bibnamefont {Mitchell}}, \bibinfo {author} {\bibfnamefont {K.}~\bibnamefont
  {Mehlawat}}, \bibinfo {author} {\bibfnamefont {Y.}~\bibnamefont {Singh}},
  \bibinfo {author} {\bibfnamefont {Y.}~\bibnamefont {Choi}}, \bibinfo {author}
  {\bibfnamefont {T.}~\bibnamefont {Gog}}, \bibinfo {author} {\bibfnamefont
  {A.}~\bibnamefont {Al-Zein}}, \bibinfo {author} {\bibfnamefont {M.~M.}\
  \bibnamefont {Sala}}, \bibinfo {author} {\bibfnamefont {M.}~\bibnamefont
  {Krisch}}, \bibinfo {author} {\bibfnamefont {J.}~\bibnamefont {Chaloupka}},
  \bibinfo {author} {\bibfnamefont {G.}~\bibnamefont {Jackeli}}, \bibinfo
  {author} {\bibfnamefont {G.}~\bibnamefont {Khaliullin}}, \ and\ \bibinfo
  {author} {\bibfnamefont {B.~J.}\ \bibnamefont {Kim}},\ }\href {\doibase
  10.1038/nphys3322} {\bibfield  {journal} {\bibinfo  {journal} {Nat. Phys.}\
  }\textbf {\bibinfo {volume} {11}},\ \bibinfo {pages} {462} (\bibinfo {year}
  {2015})}\BibitemShut {NoStop}%
\bibitem [{\citenamefont {Kim}\ \emph {et~al.}(2016)\citenamefont {Kim},
  \citenamefont {Sung}, \citenamefont {Denlinger},\ and\ \citenamefont
  {Kim}}]{kim_observation_2016}%
  \BibitemOpen
  \bibfield  {author} {\bibinfo {author} {\bibfnamefont {Y.~K.}\ \bibnamefont
  {Kim}}, \bibinfo {author} {\bibfnamefont {N.~H.}\ \bibnamefont {Sung}},
  \bibinfo {author} {\bibfnamefont {J.~D.}\ \bibnamefont {Denlinger}}, \ and\
  \bibinfo {author} {\bibfnamefont {B.~J.}\ \bibnamefont {Kim}},\ }\href
  {\doibase 10.1038/nphys3503} {\bibfield  {journal} {\bibinfo  {journal} {Nat.
  Phys.}\ }\textbf {\bibinfo {volume} {12}},\ \bibinfo {pages} {37} (\bibinfo
  {year} {2016})}\BibitemShut {NoStop}%
\bibitem [{\citenamefont {Kim}\ \emph {et~al.}(2009)\citenamefont {Kim},
  \citenamefont {Ohsumi}, \citenamefont {Komesu}, \citenamefont {Sakai},
  \citenamefont {Morita}, \citenamefont {Takagi},\ and\ \citenamefont
  {Arima}}]{kim_phase-sensitive_2009}%
  \BibitemOpen
  \bibfield  {author} {\bibinfo {author} {\bibfnamefont {B.~J.}\ \bibnamefont
  {Kim}}, \bibinfo {author} {\bibfnamefont {H.}~\bibnamefont {Ohsumi}},
  \bibinfo {author} {\bibfnamefont {T.}~\bibnamefont {Komesu}}, \bibinfo
  {author} {\bibfnamefont {S.}~\bibnamefont {Sakai}}, \bibinfo {author}
  {\bibfnamefont {T.}~\bibnamefont {Morita}}, \bibinfo {author} {\bibfnamefont
  {H.}~\bibnamefont {Takagi}}, \ and\ \bibinfo {author} {\bibfnamefont
  {T.}~\bibnamefont {Arima}},\ }\href {\doibase 10.1126/science.1167106}
  {\bibfield  {journal} {\bibinfo  {journal} {Science}\ }\textbf {\bibinfo
  {volume} {323}},\ \bibinfo {pages} {1329} (\bibinfo {year}
  {2009})}\BibitemShut {NoStop}%
\bibitem [{\citenamefont {Chen}\ \emph {et~al.}(2015)\citenamefont {Chen},
  \citenamefont {Millis},\ and\ \citenamefont
  {Marianetti}}]{chen_density_2015}%
  \BibitemOpen
  \bibfield  {author} {\bibinfo {author} {\bibfnamefont {J.}~\bibnamefont
  {Chen}}, \bibinfo {author} {\bibfnamefont {A.~J.}\ \bibnamefont {Millis}}, \
  and\ \bibinfo {author} {\bibfnamefont {C.~A.}\ \bibnamefont {Marianetti}},\
  }\href {\doibase 10.1103/PhysRevB.91.241111} {\bibfield  {journal} {\bibinfo
  {journal} {Phys. Rev. B}\ }\textbf {\bibinfo {volume} {91}},\ \bibinfo
  {pages} {241111} (\bibinfo {year} {2015})}\BibitemShut {NoStop}%
\bibitem [{\citenamefont {Park}\ \emph {et~al.}(2015)\citenamefont {Park},
  \citenamefont {Millis},\ and\ \citenamefont
  {Marianetti}}]{park_density_2015}%
  \BibitemOpen
  \bibfield  {author} {\bibinfo {author} {\bibfnamefont {H.}~\bibnamefont
  {Park}}, \bibinfo {author} {\bibfnamefont {A.~J.}\ \bibnamefont {Millis}}, \
  and\ \bibinfo {author} {\bibfnamefont {C.~A.}\ \bibnamefont {Marianetti}},\
  }\href {\doibase 10.1103/PhysRevB.92.035146} {\bibfield  {journal} {\bibinfo
  {journal} {Phys. Rev. B}\ }\textbf {\bibinfo {volume} {92}},\ \bibinfo
  {pages} {035146} (\bibinfo {year} {2015})}\BibitemShut {NoStop}%
\bibitem [{\citenamefont {Chen}\ and\ \citenamefont
  {Millis}(2016)}]{chen_spin-density_2016}%
  \BibitemOpen
  \bibfield  {author} {\bibinfo {author} {\bibfnamefont {H.}~\bibnamefont
  {Chen}}\ and\ \bibinfo {author} {\bibfnamefont {A.~J.}\ \bibnamefont
  {Millis}},\ }\href {\doibase 10.1103/PhysRevB.93.045133} {\bibfield
  {journal} {\bibinfo  {journal} {Phys. Rev. B}\ }\textbf {\bibinfo {volume}
  {93}},\ \bibinfo {pages} {045133} (\bibinfo {year} {2016})}\BibitemShut
  {NoStop}%
\bibitem [{\citenamefont {Ryee}\ and\ \citenamefont
  {Han}(2018{\natexlab{a}})}]{ryee_effect_2018}%
  \BibitemOpen
  \bibfield  {author} {\bibinfo {author} {\bibfnamefont {S.}~\bibnamefont
  {Ryee}}\ and\ \bibinfo {author} {\bibfnamefont {M.~J.}\ \bibnamefont {Han}},\
  }\href {\doibase 10.1038/s41598-018-27731-4} {\bibfield  {journal} {\bibinfo
  {journal} {Sci. Rep.}\ }\textbf {\bibinfo {volume} {8}},\ \bibinfo {pages}
  {9559} (\bibinfo {year} {2018}{\natexlab{a}})}\BibitemShut {NoStop}%
\bibitem [{\citenamefont {Ryee}\ and\ \citenamefont
  {Han}(2018{\natexlab{b}})}]{ryee_comparative_2018}%
  \BibitemOpen
  \bibfield  {author} {\bibinfo {author} {\bibfnamefont {S.}~\bibnamefont
  {Ryee}}\ and\ \bibinfo {author} {\bibfnamefont {M.~J.}\ \bibnamefont {Han}},\
  }\href {\doibase 10.1088/1361-648X/aac79c} {\bibfield  {journal} {\bibinfo
  {journal} {J. Phys.: Condens. Matter}\ }\textbf {\bibinfo {volume} {30}},\
  \bibinfo {pages} {275802} (\bibinfo {year} {2018}{\natexlab{b}})}\BibitemShut
  {NoStop}%
\bibitem [{com()}]{comment}%
  \BibitemOpen
  \href@noop {} {}\bibinfo {note} {We hereafter clearly distinguish LDA and GGA
  from LSDA and SGGA.}\BibitemShut {Stop}%
\bibitem [{\citenamefont {Dudarev}\ \emph {et~al.}(1998)\citenamefont
  {Dudarev}, \citenamefont {Botton}, \citenamefont {Savrasov}, \citenamefont
  {Humphreys},\ and\ \citenamefont
  {Sutton}}]{dudarev_electron-energy-loss_1998}%
  \BibitemOpen
  \bibfield  {author} {\bibinfo {author} {\bibfnamefont {S.~L.}\ \bibnamefont
  {Dudarev}}, \bibinfo {author} {\bibfnamefont {G.~A.}\ \bibnamefont {Botton}},
  \bibinfo {author} {\bibfnamefont {S.~Y.}\ \bibnamefont {Savrasov}}, \bibinfo
  {author} {\bibfnamefont {C.~J.}\ \bibnamefont {Humphreys}}, \ and\ \bibinfo
  {author} {\bibfnamefont {A.~P.}\ \bibnamefont {Sutton}},\ }\href {\doibase
  10.1103/PhysRevB.57.1505} {\bibfield  {journal} {\bibinfo  {journal} {Phys.
  Rev. B}\ }\textbf {\bibinfo {volume} {57}},\ \bibinfo {pages} {1505}
  (\bibinfo {year} {1998})}\BibitemShut {NoStop}%
\bibitem [{\citenamefont {Han}\ \emph {et~al.}(2006)\citenamefont {Han},
  \citenamefont {Ozaki},\ and\ \citenamefont {Yu}}]{han_$mathrmon$_2006}%
  \BibitemOpen
  \bibfield  {author} {\bibinfo {author} {\bibfnamefont {M.~J.}\ \bibnamefont
  {Han}}, \bibinfo {author} {\bibfnamefont {T.}~\bibnamefont {Ozaki}}, \ and\
  \bibinfo {author} {\bibfnamefont {J.}~\bibnamefont {Yu}},\ }\href {\doibase
  10.1103/PhysRevB.73.045110} {\bibfield  {journal} {\bibinfo  {journal} {Phys.
  Rev. B}\ }\textbf {\bibinfo {volume} {73}},\ \bibinfo {pages} {045110}
  (\bibinfo {year} {2006})}\BibitemShut {NoStop}%
\bibitem [{\citenamefont {Ozaki}(2003)}]{ozaki_variationally_2003}%
  \BibitemOpen
  \bibfield  {author} {\bibinfo {author} {\bibfnamefont {T.}~\bibnamefont
  {Ozaki}},\ }\href {\doibase 10.1103/PhysRevB.67.155108} {\bibfield  {journal}
  {\bibinfo  {journal} {Phys. Rev. B}\ }\textbf {\bibinfo {volume} {67}},\
  \bibinfo {pages} {155108} (\bibinfo {year} {2003})}\BibitemShut {NoStop}%
\bibitem [{\citenamefont {Anisimov}\ \emph {et~al.}(1993)\citenamefont
  {Anisimov}, \citenamefont {Solovyev}, \citenamefont {Korotin}, \citenamefont
  {Czyżyk},\ and\ \citenamefont
  {Sawatzky}}]{anisimov_density-functional_1993}%
  \BibitemOpen
  \bibfield  {author} {\bibinfo {author} {\bibfnamefont {V.~I.}\ \bibnamefont
  {Anisimov}}, \bibinfo {author} {\bibfnamefont {I.~V.}\ \bibnamefont
  {Solovyev}}, \bibinfo {author} {\bibfnamefont {M.~A.}\ \bibnamefont
  {Korotin}}, \bibinfo {author} {\bibfnamefont {M.~T.}\ \bibnamefont
  {Czyżyk}}, \ and\ \bibinfo {author} {\bibfnamefont {G.~A.}\ \bibnamefont
  {Sawatzky}},\ }\href {\doibase 10.1103/PhysRevB.48.16929} {\bibfield
  {journal} {\bibinfo  {journal} {Phys. Rev. B}\ }\textbf {\bibinfo {volume}
  {48}},\ \bibinfo {pages} {16929} (\bibinfo {year} {1993})}\BibitemShut
  {NoStop}%
\bibitem [{\citenamefont {Solovyev}\ \emph {et~al.}(1994)\citenamefont
  {Solovyev}, \citenamefont {Dederichs},\ and\ \citenamefont
  {Anisimov}}]{solovyev_corrected_1994}%
  \BibitemOpen
  \bibfield  {author} {\bibinfo {author} {\bibfnamefont {I.~V.}\ \bibnamefont
  {Solovyev}}, \bibinfo {author} {\bibfnamefont {P.~H.}\ \bibnamefont
  {Dederichs}}, \ and\ \bibinfo {author} {\bibfnamefont {V.~I.}\ \bibnamefont
  {Anisimov}},\ }\href {\doibase 10.1103/PhysRevB.50.16861} {\bibfield
  {journal} {\bibinfo  {journal} {Phys. Rev. B}\ }\textbf {\bibinfo {volume}
  {50}},\ \bibinfo {pages} {16861} (\bibinfo {year} {1994})}\BibitemShut
  {NoStop}%
\bibitem [{\citenamefont {Czyżyk}\ and\ \citenamefont
  {Sawatzky}(1994)}]{czyzyk_local-density_1994}%
  \BibitemOpen
  \bibfield  {author} {\bibinfo {author} {\bibfnamefont {M.~T.}\ \bibnamefont
  {Czyżyk}}\ and\ \bibinfo {author} {\bibfnamefont {G.~A.}\ \bibnamefont
  {Sawatzky}},\ }\href {\doibase 10.1103/PhysRevB.49.14211} {\bibfield
  {journal} {\bibinfo  {journal} {Phys. Rev. B}\ }\textbf {\bibinfo {volume}
  {49}},\ \bibinfo {pages} {14211} (\bibinfo {year} {1994})}\BibitemShut
  {NoStop}%
\bibitem [{\citenamefont {Liechtenstein}\ \emph {et~al.}(1995)\citenamefont
  {Liechtenstein}, \citenamefont {Anisimov},\ and\ \citenamefont
  {Zaanen}}]{liechtenstein_density-functional_1995}%
  \BibitemOpen
  \bibfield  {author} {\bibinfo {author} {\bibfnamefont {A.~I.}\ \bibnamefont
  {Liechtenstein}}, \bibinfo {author} {\bibfnamefont {V.~I.}\ \bibnamefont
  {Anisimov}}, \ and\ \bibinfo {author} {\bibfnamefont {J.}~\bibnamefont
  {Zaanen}},\ }\href {\doibase 10.1103/PhysRevB.52.R5467} {\bibfield  {journal}
  {\bibinfo  {journal} {Phys. Rev. B}\ }\textbf {\bibinfo {volume} {52}},\
  \bibinfo {pages} {R5467} (\bibinfo {year} {1995})}\BibitemShut {NoStop}%
\bibitem [{\citenamefont {MacDonald}(1983)}]{macdonald_spin-polarised_1983}%
  \BibitemOpen
  \bibfield  {author} {\bibinfo {author} {\bibfnamefont {A.~H.}\ \bibnamefont
  {MacDonald}},\ }\href {\doibase 10.1088/0022-3719/16/20/012} {\bibfield
  {journal} {\bibinfo  {journal} {J. Phys. C: Solid State Phys.}\ }\textbf
  {\bibinfo {volume} {16}},\ \bibinfo {pages} {3869} (\bibinfo {year}
  {1983})}\BibitemShut {NoStop}%
\bibitem [{\citenamefont {Bachelet}\ \emph {et~al.}(1982)\citenamefont
  {Bachelet}, \citenamefont {Hamann},\ and\ \citenamefont
  {Schlüter}}]{bachelet_pseudopotentials_1982}%
  \BibitemOpen
  \bibfield  {author} {\bibinfo {author} {\bibfnamefont {G.~B.}\ \bibnamefont
  {Bachelet}}, \bibinfo {author} {\bibfnamefont {D.~R.}\ \bibnamefont
  {Hamann}}, \ and\ \bibinfo {author} {\bibfnamefont {M.}~\bibnamefont
  {Schlüter}},\ }\href {\doibase 10.1103/PhysRevB.26.4199} {\bibfield
  {journal} {\bibinfo  {journal} {Phys. Rev. B}\ }\textbf {\bibinfo {volume}
  {26}},\ \bibinfo {pages} {4199} (\bibinfo {year} {1982})}\BibitemShut
  {NoStop}%
\bibitem [{\citenamefont {Theurich}\ and\ \citenamefont
  {Hill}(2001)}]{theurich_self-consistent_2001}%
  \BibitemOpen
  \bibfield  {author} {\bibinfo {author} {\bibfnamefont {G.}~\bibnamefont
  {Theurich}}\ and\ \bibinfo {author} {\bibfnamefont {N.~A.}\ \bibnamefont
  {Hill}},\ }\href {\doibase 10.1103/PhysRevB.64.073106} {\bibfield  {journal}
  {\bibinfo  {journal} {Phys. Rev. B}\ }\textbf {\bibinfo {volume} {64}},\
  \bibinfo {pages} {073106} (\bibinfo {year} {2001})}\BibitemShut {NoStop}%
\bibitem [{\citenamefont {Şaşıoğlu}\ \emph {et~al.}(2011)\citenamefont
  {Şaşıoğlu}, \citenamefont {Friedrich},\ and\ \citenamefont
  {Blügel}}]{sasioglu_effective_2011}%
  \BibitemOpen
  \bibfield  {author} {\bibinfo {author} {\bibfnamefont {E.}~\bibnamefont
  {Şaşıoğlu}}, \bibinfo {author} {\bibfnamefont {C.}~\bibnamefont
  {Friedrich}}, \ and\ \bibinfo {author} {\bibfnamefont {S.}~\bibnamefont
  {Blügel}},\ }\href {\doibase 10.1103/PhysRevB.83.121101} {\bibfield
  {journal} {\bibinfo  {journal} {Phys. Rev. B}\ }\textbf {\bibinfo {volume}
  {83}},\ \bibinfo {pages} {121101} (\bibinfo {year} {2011})}\BibitemShut
  {NoStop}%
\bibitem [{\citenamefont {Kim}\ \emph {et~al.}(2008)\citenamefont {Kim},
  \citenamefont {Jin}, \citenamefont {Moon}, \citenamefont {Kim}, \citenamefont
  {Park}, \citenamefont {Leem}, \citenamefont {Yu}, \citenamefont {Noh},
  \citenamefont {Kim}, \citenamefont {Oh}, \citenamefont {Park}, \citenamefont
  {Durairaj}, \citenamefont {Cao},\ and\ \citenamefont
  {Rotenberg}}]{kim_novel_2008}%
  \BibitemOpen
  \bibfield  {author} {\bibinfo {author} {\bibfnamefont {B.~J.}\ \bibnamefont
  {Kim}}, \bibinfo {author} {\bibfnamefont {H.}~\bibnamefont {Jin}}, \bibinfo
  {author} {\bibfnamefont {S.~J.}\ \bibnamefont {Moon}}, \bibinfo {author}
  {\bibfnamefont {J.-Y.}\ \bibnamefont {Kim}}, \bibinfo {author} {\bibfnamefont
  {B.-G.}\ \bibnamefont {Park}}, \bibinfo {author} {\bibfnamefont {C.~S.}\
  \bibnamefont {Leem}}, \bibinfo {author} {\bibfnamefont {J.}~\bibnamefont
  {Yu}}, \bibinfo {author} {\bibfnamefont {T.~W.}\ \bibnamefont {Noh}},
  \bibinfo {author} {\bibfnamefont {C.}~\bibnamefont {Kim}}, \bibinfo {author}
  {\bibfnamefont {S.-J.}\ \bibnamefont {Oh}}, \bibinfo {author} {\bibfnamefont
  {J.-H.}\ \bibnamefont {Park}}, \bibinfo {author} {\bibfnamefont
  {V.}~\bibnamefont {Durairaj}}, \bibinfo {author} {\bibfnamefont
  {G.}~\bibnamefont {Cao}}, \ and\ \bibinfo {author} {\bibfnamefont
  {E.}~\bibnamefont {Rotenberg}},\ }\href {\doibase
  10.1103/PhysRevLett.101.076402} {\bibfield  {journal} {\bibinfo  {journal}
  {Phys. Rev. Lett.}\ }\textbf {\bibinfo {volume} {101}},\ \bibinfo {pages}
  {076402} (\bibinfo {year} {2008})}\BibitemShut {NoStop}%
\bibitem [{\citenamefont {François}\ \emph {et~al.}(1991)\citenamefont
  {François}, \citenamefont {Lengauer}, \citenamefont {Yvon}, \citenamefont
  {Ben Yaich-Aerrache}, \citenamefont {Gougeon}, \citenamefont {Potel},\ and\
  \citenamefont {Sergent}}]{francois_structural_1991}%
  \BibitemOpen
  \bibfield  {author} {\bibinfo {author} {\bibfnamefont {M.}~\bibnamefont
  {François}}, \bibinfo {author} {\bibfnamefont {W.}~\bibnamefont {Lengauer}},
  \bibinfo {author} {\bibfnamefont {K.}~\bibnamefont {Yvon}}, \bibinfo {author}
  {\bibfnamefont {H.}~\bibnamefont {Ben Yaich-Aerrache}}, \bibinfo {author}
  {\bibfnamefont {P.}~\bibnamefont {Gougeon}}, \bibinfo {author} {\bibfnamefont
  {M.}~\bibnamefont {Potel}}, \ and\ \bibinfo {author} {\bibfnamefont
  {M.}~\bibnamefont {Sergent}},\ }\href {\doibase
  10.1524/zkri.1991.196.1-4.111} {\bibfield  {journal} {\bibinfo  {journal} {Z.
  Kristallogr. Cryst. Mater.}\ }\textbf {\bibinfo {volume} {196}},\ \bibinfo
  {pages} {111} (\bibinfo {year} {1991})}\BibitemShut {NoStop}%
\bibitem [{\citenamefont {Johrendt}(1998)}]{johrendt_crystal_1998}%
  \BibitemOpen
  \bibfield  {author} {\bibinfo {author} {\bibfnamefont {D.}~\bibnamefont
  {Johrendt}},\ }\href {\doibase
  10.1002/(SICI)1521-3749(199806)624:6<952::AID-ZAAC952>3.0.CO;2-L} {\bibfield
  {journal} {\bibinfo  {journal} {Z. Anorg. Allg. Chem.}\ }\textbf {\bibinfo
  {volume} {624}},\ \bibinfo {pages} {952} (\bibinfo {year}
  {1998})}\BibitemShut {NoStop}%
\bibitem [{\citenamefont {Pocha}\ \emph {et~al.}(2000)\citenamefont {Pocha},
  \citenamefont {Johrendt},\ and\ \citenamefont
  {Pöttgen}}]{pocha_electronic_2000}%
  \BibitemOpen
  \bibfield  {author} {\bibinfo {author} {\bibfnamefont {R.}~\bibnamefont
  {Pocha}}, \bibinfo {author} {\bibfnamefont {D.}~\bibnamefont {Johrendt}}, \
  and\ \bibinfo {author} {\bibfnamefont {R.}~\bibnamefont {Pöttgen}},\ }\href
  {\doibase 10.1021/cm001099b} {\bibfield  {journal} {\bibinfo  {journal}
  {Chem. Mater.}\ }\textbf {\bibinfo {volume} {12}},\ \bibinfo {pages} {2882}
  (\bibinfo {year} {2000})}\BibitemShut {NoStop}%
\bibitem [{\citenamefont {Nakamura}\ \emph {et~al.}(2005)\citenamefont
  {Nakamura}, \citenamefont {Chudo},\ and\ \citenamefont
  {Shiga}}]{nakamura_structural_2005}%
  \BibitemOpen
  \bibfield  {author} {\bibinfo {author} {\bibfnamefont {H.}~\bibnamefont
  {Nakamura}}, \bibinfo {author} {\bibfnamefont {H.}~\bibnamefont {Chudo}}, \
  and\ \bibinfo {author} {\bibfnamefont {M.}~\bibnamefont {Shiga}},\ }\href
  {\doibase 10.1088/0953-8984/17/38/007} {\bibfield  {journal} {\bibinfo
  {journal} {J. Phys.: Condens. Matter}\ }\textbf {\bibinfo {volume} {17}},\
  \bibinfo {pages} {6015} (\bibinfo {year} {2005})}\BibitemShut {NoStop}%
\bibitem [{\citenamefont {Guiot}\ \emph {et~al.}(2011)\citenamefont {Guiot},
  \citenamefont {Janod}, \citenamefont {Corraze},\ and\ \citenamefont
  {Cario}}]{guiot_control_2011}%
  \BibitemOpen
  \bibfield  {author} {\bibinfo {author} {\bibfnamefont {V.}~\bibnamefont
  {Guiot}}, \bibinfo {author} {\bibfnamefont {E.}~\bibnamefont {Janod}},
  \bibinfo {author} {\bibfnamefont {B.}~\bibnamefont {Corraze}}, \ and\
  \bibinfo {author} {\bibfnamefont {L.}~\bibnamefont {Cario}},\ }\href
  {\doibase 10.1021/cm200266n} {\bibfield  {journal} {\bibinfo  {journal}
  {Chem. Mater.}\ }\textbf {\bibinfo {volume} {23}},\ \bibinfo {pages} {2611}
  (\bibinfo {year} {2011})}\BibitemShut {NoStop}%
\bibitem [{\citenamefont {Ta~Phuoc}\ \emph {et~al.}(2013)\citenamefont
  {Ta~Phuoc}, \citenamefont {Vaju}, \citenamefont {Corraze}, \citenamefont
  {Sopracase}, \citenamefont {Perucchi}, \citenamefont {Marini}, \citenamefont
  {Postorino}, \citenamefont {Chligui}, \citenamefont {Lupi}, \citenamefont
  {Janod},\ and\ \citenamefont {Cario}}]{ta_phuoc_optical_2013}%
  \BibitemOpen
  \bibfield  {author} {\bibinfo {author} {\bibfnamefont {V.}~\bibnamefont
  {Ta~Phuoc}}, \bibinfo {author} {\bibfnamefont {C.}~\bibnamefont {Vaju}},
  \bibinfo {author} {\bibfnamefont {B.}~\bibnamefont {Corraze}}, \bibinfo
  {author} {\bibfnamefont {R.}~\bibnamefont {Sopracase}}, \bibinfo {author}
  {\bibfnamefont {A.}~\bibnamefont {Perucchi}}, \bibinfo {author}
  {\bibfnamefont {C.}~\bibnamefont {Marini}}, \bibinfo {author} {\bibfnamefont
  {P.}~\bibnamefont {Postorino}}, \bibinfo {author} {\bibfnamefont
  {M.}~\bibnamefont {Chligui}}, \bibinfo {author} {\bibfnamefont
  {S.}~\bibnamefont {Lupi}}, \bibinfo {author} {\bibfnamefont {E.}~\bibnamefont
  {Janod}}, \ and\ \bibinfo {author} {\bibfnamefont {L.}~\bibnamefont
  {Cario}},\ }\href {\doibase 10.1103/PhysRevLett.110.037401} {\bibfield
  {journal} {\bibinfo  {journal} {Phys. Rev. Lett.}\ }\textbf {\bibinfo
  {volume} {110}},\ \bibinfo {pages} {037401} (\bibinfo {year}
  {2013})}\BibitemShut {NoStop}%
\bibitem [{\citenamefont {Camjayi}\ \emph {et~al.}(2014)\citenamefont
  {Camjayi}, \citenamefont {Acha}, \citenamefont {Weht}, \citenamefont
  {Rodríguez}, \citenamefont {Corraze}, \citenamefont {Janod}, \citenamefont
  {Cario},\ and\ \citenamefont {Rozenberg}}]{camjayi_first-order_2014}%
  \BibitemOpen
  \bibfield  {author} {\bibinfo {author} {\bibfnamefont {A.}~\bibnamefont
  {Camjayi}}, \bibinfo {author} {\bibfnamefont {C.}~\bibnamefont {Acha}},
  \bibinfo {author} {\bibfnamefont {R.}~\bibnamefont {Weht}}, \bibinfo {author}
  {\bibfnamefont {M.~G.}\ \bibnamefont {Rodríguez}}, \bibinfo {author}
  {\bibfnamefont {B.}~\bibnamefont {Corraze}}, \bibinfo {author} {\bibfnamefont
  {E.}~\bibnamefont {Janod}}, \bibinfo {author} {\bibfnamefont
  {L.}~\bibnamefont {Cario}}, \ and\ \bibinfo {author} {\bibfnamefont {M.~J.}\
  \bibnamefont {Rozenberg}},\ }\href {\doibase 10.1103/PhysRevLett.113.086404}
  {\bibfield  {journal} {\bibinfo  {journal} {Phys. Rev. Lett.}\ }\textbf
  {\bibinfo {volume} {113}},\ \bibinfo {pages} {086404} (\bibinfo {year}
  {2014})}\BibitemShut {NoStop}%
\bibitem [{\citenamefont {Arita}\ \emph {et~al.}(2012)\citenamefont {Arita},
  \citenamefont {Kuneš}, \citenamefont {Kozhevnikov}, \citenamefont
  {Eguiluz},\ and\ \citenamefont {Imada}}]{arita_ab_2012}%
  \BibitemOpen
  \bibfield  {author} {\bibinfo {author} {\bibfnamefont {R.}~\bibnamefont
  {Arita}}, \bibinfo {author} {\bibfnamefont {J.}~\bibnamefont {Kuneš}},
  \bibinfo {author} {\bibfnamefont {A.~V.}\ \bibnamefont {Kozhevnikov}},
  \bibinfo {author} {\bibfnamefont {A.~G.}\ \bibnamefont {Eguiluz}}, \ and\
  \bibinfo {author} {\bibfnamefont {M.}~\bibnamefont {Imada}},\ }\href
  {\doibase 10.1103/PhysRevLett.108.086403} {\bibfield  {journal} {\bibinfo
  {journal} {Phys. Rev. Lett.}\ }\textbf {\bibinfo {volume} {108}},\ \bibinfo
  {pages} {086403} (\bibinfo {year} {2012})}\BibitemShut {NoStop}%
\end{thebibliography}%

\end{document}